\documentclass[twocolumn,prb,showpacs,amsmath,amssymb,superscriptaddress]{revtex4}

\pdfoutput=1

\usepackage{xcolor}
\usepackage{times}
\usepackage{amsfonts}
\usepackage{amssymb}
\usepackage{amsmath}
\usepackage{graphicx}
\usepackage{bm}
\usepackage{verbatim}

\usepackage{hyperref}

\usepackage{bm} 
\usepackage{color}
\usepackage{stackrel}
\usepackage{accents}

\usepackage{latexsym}

\newcommand{\bsub}{\begin{subequations}}
\newcommand{\esub}{\end{subequations}}

\newcommand \bea {\begin{eqnarray} }
\newcommand \eea {\end{eqnarray}}
 
\newcommand{\beg}{\begin{equation}}
\newcommand{\en}{\end{equation}}
\newcommand{\bp}{\mathbf p}

\newcommand{\br}{\mathbf r}

\newcommand \bel  {\begin{align}}
\newcommand \enl  {\end{align}}

\newcommand{\dg}{^\dagger}

\newcommand{\pmat}{\begin{pmatrix}}
\newcommand{\epmat}{\end{pmatrix}}

\def\8{\infty}

\def\undertext#1{\vtop{\hbox{#1}\kern 1pt \hrule}}

\def\be{\begin{equation}}
\def\ee{\end{equation}}
\def\bea{\begin{eqnarray} & &}
\def\eea{\end{eqnarray}}

\begin{document}

\title{Gaussian fluctuation corrections to a mean-field theory of complex hidden order in URu$_2$Si$_2$}

\author{Pengtao Shen}
\affiliation{Department of Physics, Kent State University, Kent, OH 44242, USA}

\author{Maxim Dzero}
\affiliation{Department of Physics, Kent State University, Kent, OH 44242, USA}
\affiliation{Max Planck Institute for the Physics of Complex Systems, N\"{o}thnitzer str. 38, 01187 Dresden, Germany}

\begin{abstract} 
Hidden-order phase transition in the heavy-fermion superconductor URu$_2$Si$_2$ exhibits the mean-field-like anomaly in temperature dependence of heat capacity. Motivated by this observation, here we explore the impact of the complex order parameter fluctuations on the thermodynamic properties of the hidden order phase. Specifically, we employ the mean-field theory for the hidden order which describes the hidden order parameter by an average of the hexadecapole operator. We compute the gaussian fluctuation corrections to the mean-field theory equations including both the fluctuations due to 'hidden order' as well as antiferromagnetic order parameters. We find that the gaussian fluctuations lead to the smearing of the second-order transition rendering it to become the first-order one. The strength of the first-order transition is weakly dependent on the strength of underlying antiferromagnetic exchange interactions.
\end{abstract}


\maketitle

\section{Introduction}
Virtually every  textbook on thermodynamics and statistical physics includes the discussion of the first and second order phase transitions. The latter are defined by the discontinuity of the second derivative of free energy at some critical temperature $T_c$. The defining feature of the second-order phase transitions is that the low temperature ordered phase has lower symmetry than the high-temperature disordered one and one can conveniently introduce the order parameter to describe the transition.\cite{LL} Therefore, it is usually possible to associate the physical observable (and the corresponding susceptibility) with the order parameter by identifying what symmetry has been broken by transitioning into the ordered phase: for example, the time-reversal symmetry corresponds to a state with finite magnetization while breaking of the global $U(1)$ gauge symmetry signals an onset of superconductivity. 

Given the remarkable success in our understanding of the phase transitions and advances in experimental techniques, a relative ease with which one can identify the symmetry of the low-temperature state is almost always taken for granted.
The intriguing exception to this state of affairs was furnished by the observation of the second-order phase transition in URu$_2$Si$_2$ at temperature $T_c\approx 17.5$ K.\cite{HOMydosh1985} 
Indeed, despite more than thirty years of intensive theoretical and experimental research, the consensus on the nature of the broken symmetry state has not been reached yet (for details on competing theories of hiddent order and recent experimental efforts we would like to refer the reader to an excellent recent review paper by Mydosh [\onlinecite{Review2014}] and references therein).

Experimentally, one of the intriguing features of the hidden-order phase transition is a sharp  -- mean-field-like -- discontinuity in the temperature dependence of the heat capacity. If we were to entertain an idea that the hidden order phase transition is governed by the itinerant degrees of freedom, we would find that the corresponding 'hidden-order' susceptibility must become logarithmically divergent with temperature. Then an analogy with the conventional superconductivity immediately comes to mind for the superconducting transition in elemental metals has also sharp jump in the heat capacity at the critical temperature and, consequently, the superconductivity is fairly accurately described by the BCS mean-field theory.\cite{BCS} In fact, Kos, Millis and Larkin have demonstrated that although the gaussian fluctuation corrections to the BCS mean-field equations due to the amplitude and phase fluctuations are logarithmically diverging, the corresponding divergences cancel each other out in the mean-field equations \cite{Kos2004} making the BCS mean-field approximation very accurate (see also Ref. [\onlinecite{Joerg1}] for a related discussion). Interestingly, most recently Hoyer and Schmalian have shown that similar cancellation of the logarithms does not happen for the case of the charge-density-wave transition (CDW) and spin-density-wave (SDW) transition in one or three spacial dimensions, which naturally renders the mean-field treatment of those transitions uncontrolled.\cite{Joerg2} The only exception is the SDW transition in two dimensions with the \emph{perfectly nested} Fermi surface. \cite{Joerg2} In view of these theoretical considerations along with the recent experimental results,\cite{Review2014} it seems perfectly reasonable to us to think that the hidden order transitions is likely driven by the \emph{local} and not itinerant degrees of freedom. 

Recenly, Haule and Kotliar (HK) have employed the combination of the density functional theory together with the dynamical mean-field theory to put forward a theory for the hidden-order transition, which is governed by the local $5f$ orbital degrees of freedom.\cite{KH2009}
The corresponding order parameter is a complex function, whose real part accounts for the hidden order phase: it is determined by the average of the hexadecapole operator and corresponds to an excitonic mixing between the two lowest lying states originating from the uranium $5f$ non-Kramers doublets which are split by the crystalline electric fields. The imaginary part of the order parameter accounts for an antiferromagnetic order which emerges when the external pressure is applied. In their follow-up paper, \cite{KH2010} Haule and Kotliar have developed a Landau-Ginzburg description of hidden order state with the complex order parameter. 

Motived by these developments, in this paper we present the results of our calculations of the gaussian fluctuation corrections to the mean-field theory equations of the complex 'hidden-order' parameter.  
We find that in the vicinity of the hidden-order transition the gaussian fluctuations of the real and imaginary parts of the complex order parameter render the transition to become first order: at the critical temperature the absolute value of the order parameter changes abruptly from zero to some finite value which is close to the mean-field value calculated at $T=0$. By calculating the dependence of the critical temperature on the value of the antiferromagnetic exchange interaction, we also find that the gaussian fluctuations have generically substantial effect on the value of the mean-field critical temperature: the critical temperature is decreased approximately by a factor of two. 

Our paper is organized as follows. In the next Section we introduce the microscopic model to describe the hidden-order transition. Section III provides the summary for the mean-field approximation of our model. The derivation of the gaussian fluctuation corrections to the mean-field equations are presented in Section IV. Finally, Section V is devoted to discussion of our results and conclusions. Throughout the paper we use the units $\hbar=k_B=1$. 

\section{Model}
Following the discussion in Ref. [\onlinecite{KH2010}], we consider uranium ion in $5f^2$ valence configuration corresponding to a state with the total angular momentum $J=4$. As a result of the crystalline electric fields, the nine-fold degeneracy is lifted. The first principles calculations \cite{KH2009} showed that the two lowest lying state is a non-Kramers doublet which can be written as a linear combination of the eigenvectors of the angular momentum operator $\hat{J}_z$: 
\beg\label{doublet}
\begin{split}
&|\gamma_0\rangle=\frac{i}{\sqrt{2}}\left(|4\rangle-|-4\rangle\right), \\ 
&|\gamma_1\rangle=\frac{\cos\phi}{\sqrt{2}}\left(|4\rangle+|-4\rangle\right)+\sin\phi|0\rangle,
\end{split}
\en
where $\phi$ is some parameter whose specific value will not be important for our subsequent discussion. Since we are considering only two states (\ref{doublet}), it is convenient to represent 
them using the fermionic creation operators $|\gamma_a\rangle=\hat{f}_a\dg|\textrm{vac}\rangle$.
We write the model Hamiltonian $\hat{H}$ in terms of the fermionic operators as follows:
\beg\label{Eq1}
\begin{split}
\hat{H}&=-\Delta_z\sum\limits_{iab}\hat{f}_{ia}\dg\sigma_{ab}^z\hat{f}_{ib}\\&-\frac{1}{2}\sum\limits_{ij,ab}\sum\limits_{\alpha=x,y}u_{ij}^{\alpha}\left(\hat{f}_{ia}\dg\sigma_{ab}^\alpha\hat{f}_{ib}\right)\left(\hat{f}_{jc}\dg\sigma_{cd}^\alpha\hat{f}_{jd}\right).
\end{split}
\en
Here the summations are performed over the lattice sites ${\vec r}_i$ and the fermionic states $a,b=0,1$, $2\Delta_z$ is the energy splitting between the states $|\gamma_0\rangle$ and $|\gamma_1\rangle$, couplings $u_{ij}^{x}$ and $u_{ij}^y$ account for the interaction between the two-level systems and $\sigma^{x}$, $\sigma^{y}$ and $\sigma^{z}$ are Pauli matrices. The exchange coupling $u_{ij}^x$ drives the hidden-order transition with the order parameter $\psi_{x}({\vec r}_i)=\langle \hat{f}_{i0}\dg\hat{f}_{i1}+\hat{f}_{i1}\dg\hat{f}_{i0}\rangle$, which is proportional to the matrix element of the hexadecapole operator
\linebreak $(\hat{J}_x\hat{J}_y+\hat{J}_y\hat{J}_x)(\hat{J}_x^2-\hat{J}_y^2)$ hence the name 'hexadecapole order'. Lastly, the exchange couplings $u_{ij}^y$ accounts for the antiferromagnetic correlations along the $z$ axis and, in principle, may lead to antiferromagnetic order described by the expectation value $\psi_{y}({\vec r}_i)=i\langle \hat{f}_{i0}\dg\hat{f}_{i1}-\hat{f}_{i1}\dg\hat{f}_{i0}\rangle\propto\langle 1|\hat{J}_z|0\rangle$. Lastly, we mention that in the case when external magnetic field is applied, the model Hamiltonian would contain an extra term $H_Z= 
-\Delta_y\sum\hat{f}_{ia}\dg\sigma_{ab}^y\hat{f}_{ib}$.

For our subsequent analysis of the model (\ref{Eq1}), it will be convenient to use the path integral formulation. Since we are essentially dealing with the spin-1/2 operators in the fermionic representation, we can use the method by Popov and Fedotov who developed the path integral formalism for spin systems.\cite{PopovFedotov1988} The constraint which excludes double occupation on each site due to the Pauli principle, $\sum_{a}\hat{f}_{ia}\dg\hat{f}_{ia}=1$, is taken into account by introducing the complex chemical potential \linebreak  $\mu_{\textrm{pf}}=-i\pi T/2$. Thus, the action for our problem reads
\beg\label{Action}
S=\int\limits_0^\beta d\tau\left\{\sum\limits_{ia}\overline{f}_{ia}\left(\frac{\partial}{\partial \tau}-\mu_{\textrm{pf}}\right){f}_{ia}+H(\overline{f},f)\right\},
\en
where $\overline{f}_{ia}(\tau)$ and $f_{ia}(\tau)$ are mutually independent Grassmann variables. The partition function is determined by 
\beg\label{Z}
{\cal Z}=\int D[\overline{f},f]e^{-S}.
\en 
We proceed by performing the Hubbard-Stratonovich transformation by introducing the local bosonic fields ${\vec \Phi}_j(\tau)$ which couple linearly to the fermionic fields ${\vec b}_i(\tau)=\sum_{ab}\overline{f}_{ia}(\tau){\vec \sigma}_{ab}f_{ib}(\tau)$: 
\beg\label{Bosonic}
\begin{split}
&e^{\frac{1}{2}\sum u_{ij}^\alpha b_{i\alpha}b_{j\alpha}}=\frac{1}{\textrm{C}}\int D{\vec{\Phi}}
e^{-\frac{1}{2}\sum\Phi_i^\alpha J_{ij}^{\alpha}\Phi_i^\alpha+\sum{\vec \Phi}_j\cdot{\vec b}_j},
\end{split}
\en
where the summations are performed over repeated indices, $J_{ij}^{\alpha}=[u^{\alpha}]_{ij}^{-1}$ and $\textrm{C}=4\pi^2\textrm{det}[u^x]\textrm{det}[u^y]$. The resulting action becomes gaussian for the fermionic fields. Performing an integration over the fermionic fields yields:
\beg\label{BosonicAction}
S[{\vec \Phi}]=\frac{1}{2}\int\limits_0^\beta d\tau\sum\Phi_i^\alpha J_{ij}^{\alpha}\Phi_i^\alpha-\textrm{Tr}\log\hat{\cal G}^{-1},
\en 
where the matrix $\hat{\cal G}^{-1}$ is defined by
\beg\label{calG}
\hat{\cal G}^{-1}({\vec r}_j,\tau)=\left(\partial_\tau-\mu_{\textrm{pf}}\right)\sigma_0-\Delta_z\sigma^z+{\vec \Phi}_j(\tau)\cdot{\vec \sigma},
\en
where $\sigma_0$ is a unit matrix. In the next Section we will analyze the action (\ref{BosonicAction}) using the mean-field theory. 

\section{Review of the mean-field theory}
The mean-field theory for the hidden order corresponds to the saddle-point approximation to Eq. (\ref{BosonicAction}):
\beg\label{SaddlePoint}
{\vec \Phi}_j(\tau)={\vec \psi}.
\en
At the saddle point the action is proportional to the free energy. Minimizing the free energy with respect to $\psi_x$ and $\psi_y$ yields the following mean-field equations:
\beg\label{mfeqs}
\begin{split}
\frac{1}{2}\sum\limits_j J_{ij}^x\psi_x&=T\sum\limits_{\nu_m}\frac{\psi_x}{(\nu_m+\frac{\pi T}{2})^2+\Delta_z^2+{\vec \psi^2}}, \\
\frac{1}{2}\sum\limits_j J_{ij}^y\psi_y&=T\sum\limits_{\nu_m}\frac{\psi_y}{(\nu_m+\frac{\pi T}{2})^2+\Delta_z^2+{\vec \psi^2}}
\end{split}
\en 
and the summation is carried over the fermionic Matsubara frequencies $\nu_m=\pi T(2m+1)$. 
For our purposes, it will suffice to consider the case when the second mean-field equation (\ref{mfeqs}) has only trivial solution, $\psi_y=0$. This implies that the ground state has purely real (hexadecapole) order parameter, $\psi_x\not=0$.  

To make further progress, some assumptions about the exchange couplings $u_{ij}^x$ must be made. 
Let us consider the simplest case of nearest neighbor interactions
\beg\label{uijx}
u_{ij}^x=U_x\delta_{j,i\pm\delta}
\en
where ${\vec r}_{i\pm\delta}$ denotes the positions of the nearest neighbors. Assuming the periodic boundary conditions, for the inverse of the matrix $J_{ij}^x$ it obtains
\beg\label{sumJx}
\sum\limits_jJ_{ij}^x=\frac{1}{2U_x}.
\en
Note that in this equations the summation extends over all lattice sites. 

Given (\ref{uijx}), we can easily perform the summations over the Matsubara frequencies,  so the equation which determines the temperature dependence of the order parameter reads
\beg\label{psixT}
\frac{\sqrt{\Delta_z^2+\psi_x^2}}{2U_x}=\tanh\left(\beta\sqrt{\Delta_z^2+\psi_x^2}\right),
\en
where $\beta=1/T$. For the critical temperature one finds
\beg\label{Tcmf}
T_c^{(\textrm{mf})}=\frac{2\Delta_z}{\log\left(\frac{2U_x+\Delta_z}{2U_x-\Delta_z}\right)},
\en
which, given the choice (\ref{uijx}), implies that the exchange coupling constant must satisfy $U_x>\Delta_z/2$. At temperatures just below the critical temperature for the order parameter we find
\beg\label{psixTc}
\psi_x(t)\approx\frac{2\Delta_z\sqrt{2\beta_c\Delta_z}}{\sqrt{{\sinh(2\beta_c\Delta_z)}{}-2\beta_c\Delta_z}}\sqrt{t}, 
\en
where $\beta_c=1/T_c^{(\textrm{mf})}$ and $t=(T_c^{(\textrm{mf})}-T)/T_c^{(\textrm{mf})}\ll 1$.
Finally, for the free energy (normalized by the number of lattice sites) we have 
\beg\label{F0}
F_0=\frac{\psi_x^2}{4U_x}-T\log\left[2\cosh(\sqrt{\Delta_z^2+\psi_x^2}/T)\right]
\en
and the heat capacity is
\beg\label{HC}
C_{\textrm{mf}}(T)=\frac{\beta^2(\Delta_z^2+\psi_x^2-T\psi_x\frac{d\psi_x}{dT})}{\cosh^2(\beta\sqrt{\Delta_z^2+\psi_x^2})}.
\en
Expression (\ref{psixTc}) allows one to evaluate the jump of the heat capacity at $T=T_c$:
\beg\label{jump}
\Delta C_{\textrm{mf}}(T_c)=\frac{2(\beta_c\Delta_z)^3}{\cosh^2(\beta_c\Delta_z)\left[
{\sinh(2\beta_c\Delta_z)}-2\beta_c\Delta_z\right]}.
\en
This result agrees with the corresponding expression in Ref. [\onlinecite{KH2010}]. It obviously allows one to obtain an estimate for the value of the parameter $\Delta_z$ as well the value of the exchange constant $U_x$.
\section{Gaussian fluctuation corrections}
The fluctuations of the order parameter $\psi_x$ as well as $\psi_y$ will modify the mean-field equations above. To study fluctuations we write
\beg\label{Fluctuate}
{\vec \Phi}_j(\tau)={\vec \psi}+{\vec \eta}_j(\tau).
\en
We insert this expression into (\ref{BosonicAction}) and then expand the resulting actions in powers of ${\vec \eta}_j(\tau)$. Since the linear in ${\vec \eta}_j(\tau)$ term vanishes, so the first non-zero term in the action will be quadratic in ${\vec \eta}$. Restricting the expansion to quadratic order, for the action it formally obtains:
\beg\label{NewAction}
S[{\vec \Phi}]=S_0[{\vec \psi}]+S_2[{\vec \eta}].
\en
Here the first term is (\ref{BosonicAction}) evaluated within the saddle-point approximation, while the remaining term describes the effect of fluctuations at the gaussian level:

\begin{figure}
\centering
\includegraphics[width=0.95\linewidth]{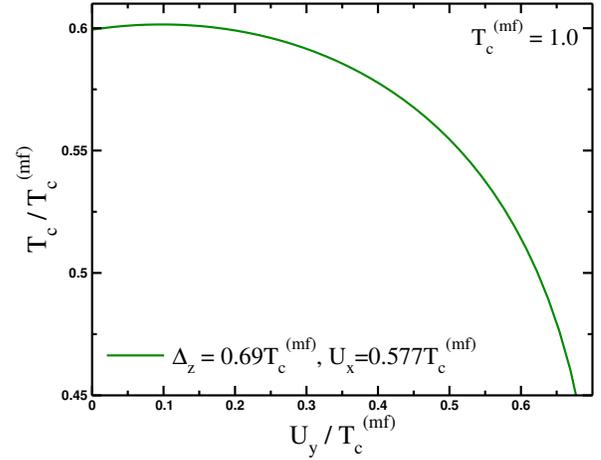}
\caption{Critical temperature $T_c$ of the hidden-order transition plotted in the units of the mean-field critical temperature $T_c^{\textrm{mf}}$ for $N_b=6$. The reduction in the values of $T_c$ is due to gaussian fluctuations which increase with an increase in the value of antiferromagnetic exchange coupling $U_y$.}
\label{Fig1Tc}
\end{figure}

\beg\label{Gaussian}
\begin{split}
S_2[\vec \eta]=\frac{1}{2}\int\limits_0^\beta d\tau
\sum\limits_{ij}\eta_{ia}(\tau)\left[J_{ij}^a\delta_{ab}+\Pi_{ab}(\tau)\right]\eta_{jb}(\tau), \\
\end{split}
\en
where 
\beg\label{Piab}
\Pi_{ab}(i\nu)=T\sum\limits_{i\omega}\textrm{Tr}\left[\hat{\cal G}_0(i\omega+i\nu)\hat{\sigma}_a\hat{\cal G}_0(i\omega)\hat{\sigma}_b\right]
\en
and $\hat{\cal G}_0(\tau)$ can be obtained from (\ref{calG}) by replacing ${\vec \Phi}_j$ with its saddle-point value.
 
Since the action (\ref{NewAction}) is gaussian, we can formally integrate out the fluctuating fields ${\vec \eta}_j$. For the fluctuation correction to the free energy we found
\beg\label{Fg}
F_2({\vec \psi})=T\textrm{Tr}\log\left[\hat{1}+\hat{J}^{-1}\cdot\hat{\Pi}\right].
\en
Here the elements of matrix $\hat{J}^{-1}$ can be conveniently written in momentum representation
\beg\label{InvJ}
\hat{J}^{-1}=\left(\begin{matrix} U_x(\bp) & 0 \\ 0 & U_y(\bp)\end{matrix}\right)
\en
and the momentum dependence of both $U_x(\bp)$ and $U_y$ is obtained from
\beg\label{Ua}
U_a(\bp)=\frac{1}{N_b}\sum\limits_j u_{0j}^a e^{i\bp\cdot \br_j}
\en
and $N_b$ equals to the number of the nearest neighbors. Note that parameter $1/N_b$ serves as a control parameter of the theory: when $N_b\to\infty$ the contribution of gaussian fluctuations to the free energy vanishes. Finally, we remind the reader that in the calculation of $U_y(\bp)$ we have to take
into account that $u_{i,i\pm\delta}^y$ changes sign for the nearest neighbors along the $z$ axis.
\begin{figure}
\centering
\includegraphics[width=0.95\linewidth]{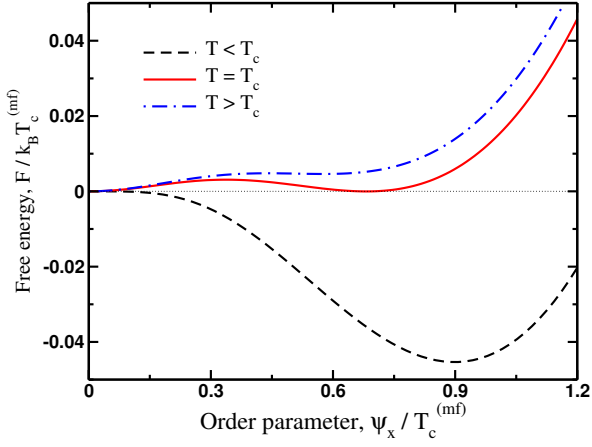}
\caption{{\small (Color online)} Free energy dependence on the value of the order parameter
calculated at various temperatures.}
\label{Fig2Free}
\end{figure}
\begin{figure}
\centering
\includegraphics[width=0.95\linewidth]{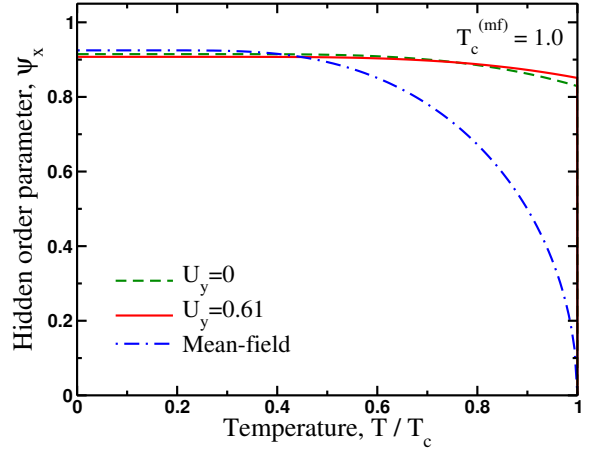}
\caption{{\small (Color online)} Temperature dependence of the order parameter $\psi_x$ in the presence of gaussian fluctuations and at the mean-field level for $N_b=6$. Gaussian fluctuations lead to a sudden
change of the order parameter at $T_c$: the second-order mean-field transition becomes the first-order transition.}
\label{Fig3Psix}
\end{figure}
Calculation of the trace in (\ref{Fg}) is straightforward:
\beg\label{F22}
F_2({\vec \psi})=2T\int\frac{d^3\bp}{(2\pi)^3}\log\left[\frac{\sinh\left(\beta R_\bp\right)}{\sinh\left(\beta\sqrt{\Delta_z^2+\psi_x^2}\right)}\right].
\en
Here the momentum integrals are performed over the first Brillouin zone and 
function $R_\bp$ is defined by
\beg\label{Rp}
\begin{split}
R_\bp^2&=\Delta_z^2+\psi_x^2-\frac{U_x(\bp)\Delta_z^2\tanh(\beta\sqrt{\Delta_z^2+\psi_x^2})}{\sqrt{\Delta_z^2+\psi_x^2}}\\&-U_y(\bp)\sqrt{\Delta_z^2+\psi_x^2}\tanh(\beta\sqrt{\Delta_z^2+\psi_x^2})\\\ &+\frac{\Delta_z^2U_x(\bp)U_y(\bp)\tanh^2(\beta\sqrt{\Delta_z^2+\psi_x^2})}{{\Delta_z^2+\psi_x^2}}.
\end{split}
\en
From this expression it is clear that despite the fact that in the hidden order state $\psi_y=0$, the antiferromagnetic fluctuations contribute the free energy.

The free energy is given by the sum of $F_0$ and $F_2$, Eqs. (\ref{F0},\ref{F22}) and we can determine the fluctuation corrections to critical temperature and order parameter. Given the momentum dependence of the functions $U_x(\bp)$ and $U_y(\bp)$, the calculation of the momentum integrals will have to be performed numerically. 
In Fig. \ref{Fig1Tc} we present the calculation of the critical temperature as a function of antiferromagnetic exchange coupling $U_y$. Perhaps not very surprisingly we find that to contribution of the gaussian fluctuations grows with an increase in $U_y$ leading to an overall suppression of $T_c$. 

In Fig. \ref{Fig2Free} we show the dependence of the free energy on the order parameter $\psi_x$
at various temperatures. The most surprising result we find that exactly at $T=T_c$ the free energy has double minimum. 
In Fig. \ref{Fig3Psix} we show the temperature dependence of the order parameter. In agreement with the free energy calculation, we find that fluctuations have a profound effect on the order parameter: they drive the mean-field transition to become the first-order one. The strength of the first order transition is obviously determined by $N_b$.

\section{Conclusions}
In this paper we have computed the gaussian fluctuation corrections to the mean-field theory of the hidden-order transition with the complex order parameter. Under an assumption of nearest neighbors interactions, we found that gaussian fluctuations drive the transition first-order. It is certainly possible that including the fluctuations beyond the gaussian approximation will make the transition weakly first order or will lead to the cancellation of the gaussian correction rendering the transition second order as manifested by experiments. However, our results clearly show that fully consistent mean-field theory of the hidden order transition still awaits its development. 

\paragraph{Acknowledgments.} The authors are grateful to Kristjan Haule, Piers Coleman, John Mydosh and Peter Riseborough for  useful discussions. We acknowledge the financial support by the National Science Foundation grant NSF-DMR-1506547.
The work of one of us (M.D.) was also financially supported in part by the U.S. Department of Energy, Office of Science, Office of Basic Energy Sciences under Award No. DE-SC0016481 and MPI-PKS (Dresden). M.D. thanks the Visitors Program at MPI-PKS for hospitality.  

\bibliography{khgauss1}

\end{document}